\documentclass[nofootinbib,aps,11pt]{revtex4-1}
\usepackage{graphicx}
\usepackage{hyperref}
\usepackage{cancel}
\usepackage{amssymb}
\usepackage{textcomp}
\usepackage{amsmath}
\usepackage{bm}
\usepackage{times}
\usepackage{epsfig}
\usepackage{color}

\begin{document}
\title{\Large Prospects of $Z'$ Portal Dark Matter in $U(1)_{L_\mu-L_\tau}$}
\bigskip
\author{Zhen-Wei Wang$^1$}
\author{Zhi-Long Han$^1$}
\email{sps\_hanzl@ujn.edu.cn}
\author{Fei Huang$^1$}
\email{sps\_huangf@ujn.edu.cn}
\author{Yi Jin$^{1,2}$}
\author{Honglei Li$^1$}
\email{sps\_lihl@ujn.edu.cn}
\affiliation{
$^1$School of Physics and Technology, University of Jinan, Jinan, Shandong 250022, China
\\
$^2$Guangxi Key Laboratory of Nuclear Physics and Nuclear Technology, Guangxi Normal University, Guilin, Guangxi 541004, China}
\date{\today}

\begin{abstract}
The gauged $U(1)_{L_\mu-L_\tau}$ model is well-motivated to explain the muon $g-2$ anomaly and dark matter in previous studies. However, the latest NA64$\mu$ experiment has almost excluded all the parameter space for the muon $g-2$, which indicates that the light dark matter benchmark scenarios interpreting muon $g-2$ in previous studies are also not allowed at present. In light of many recent and future experimental results, we revisit the minimal $Z'$ portal dark matter in $U(1)_{L_\mu-L_\tau}$. Focus on the phenomenology of dark matter $\chi$, we first explore the viable parameter space for light dark matter under various tight constraints. Around the $Z'$ resonance, we find that there is still a large parameter space for $m_\chi\gtrsim10$ MeV via thermal freeze-out. On the other hand, the constraints on $Z'$ above the electroweak scale are quite loose, but are less studied for dark matter. We also investigate the heavy dark matter around the TeV scale and corresponding constraints. A large part of the parameter space for dark matter is within the reach of future experiments.

\end{abstract}

\maketitle

\section{Introduction}

The standard model of particle physics is quite successful at various collider experiments. However, the cosmological and astrophysical observations support the existence of particle dark matter \cite{Cirelli:2024ssz}, which is absent in the standard model. The weakly interacting massive particle (WIMP), produced via the thermal freeze-out mechanism, can naturally obtain the observed dark matter relic density with electroweak scale couplings \cite{Roszkowski:2017nbc}. Due to the lack of a positive dark matter signal, the WIMP scenario might be tightly constrained by the collider and direct detection experiments \cite{Arcadi:2017kky}. It should be noted that these tight collider and direct detection limits depend heavily on the WIMP interactions with the quarks\cite{Arcadi:2024ukq}.
Therefore, one pathway to avoid these limits is assuming the leptophilic dark matter \cite{Fox:2008kb,Cohen:2009fz,Bai:2014osa,Chang:2014tea}.

Another hint of new physics beyond the standard model is the $4.2\sigma$ excess of the muon $g-2$ \cite{Muong-2:2021ojo}. This discrepancy can be resolved by introducing the new gauge boson $Z'$ in the $U(1)_{L_\mu-L_\tau}$ symmetry \cite{He:1990pn,Baek:2001kca}, which contributes to the anomalous muon magnetic moment at one-loop level \cite{Lindner:2016bgg}. The new gauge boson $Z'$ is leptophilic, thus it is also extensively considered as the mediator of dark matter \cite{Baek:2008nz,Baek:2015mna,Baek:2015fea,Patra:2016shz,Biswas:2016yjr,Kamada:2018zxi,Qi:2021rhh,Baek:2022ozm,Wang:2023pnd,Figueroa:2024tmn,Choubey:2024krp}. The common origin of muon $g-2$ and light dark matter can be realized with $g'\sim5\times10^{-4}$ and $m_{Z'}\sim\mathcal{O}(10)$ MeV \cite{Foldenauer:2018zrz,Holst:2021lzm}. However, the latest NA64$\mu$ experiment has almost excluded all the common parameter space for the muon $g-2$ and light dark matter \cite{NA64:2024klw}. So the benchmark scenarios of light dark matter in previous studies are not suitable at present. 

Since the new gauge boson $Z'$ does not directly couple to electron and quarks, the constraints for heavy $Z'$ above the electroweak scale from colliders are relatively loose. For instance, the current LHC search could only probe $Z'$ from the final state radiation of $\mu$ or $\tau$, which is only sensitive when $m_{Z'}<m_Z$ \cite{ATLAS:2023vxg,ATLAS:2024uvu}. The gauge boson $Z'$ is also lepton flavor dependent, thus it can explain the deviation of lepton universality in beauty-quark decays with heavy dark matter \cite{Altmannshofer:2016jzy,Biswas:2019twf,Han:2019diw,LHCb:2021trn,Altmannshofer:2021qrr}. Recently, the TeV energy scale muon collider has been proposed \cite{Delahaye:2019omf,Accettura:2023ked}, which is promising to probe the muon-philic $Z'$ above the electroweak scale \cite{Huang:2021nkl,Dasgupta:2023zrh,Sun:2023ylp,Korshynska:2024suh}.

In light of recent and upcoming experiments, we revisit the $Z'$ portal dark matter in the $U(1)_{L_\mu-L_\tau}$ symmetry. We focus on the phenomenology of dark matter, meanwhile excess of muon $g-2$ might be resolved by other new physics \cite{Lindner:2016bgg}. The simplest scenario of Dirac dark matter $\chi$ with the $U(1)_{L_\mu-L_\tau}$ charge $Q_\chi=1$ is illustrated in this paper. Other possible assignment of $Q_\chi$ and scalar dark matter can be found in Ref. \cite{Krnjaic:2019rsv,Asai:2020qlp,Drees:2021rsg,Hapitas:2021ilr,Deka:2022ltk}.

This paper is organized as follows. In Section \ref{Sec:MD}, we briefly review the $Z'$ portal dark matter in $U(1)_{L_\mu-L_\tau}$ and various experimental constraints. The phenomenology of dark matter is discussed in Section \ref{Sec:DM}. Finally, conclusions are presented in Section \ref{Sec:CL}.

\section{The Model}\label{Sec:MD}

For the simplest Dirac dark matter with the $U(1)_{L_\mu-L_\tau}$ charge $Q_\chi=1$, the relevant interactions of the new gauge boson $Z'$ are \cite{Holst:2021lzm}
\begin{equation}
	\mathcal{L}\supset g'(\bar{\mu}\gamma^\mu\mu-\bar{\tau}\gamma^\mu\tau+\bar{\nu}_\mu\gamma^\mu P_L\nu_\mu-\bar{\nu}_\tau\gamma^\mu P_L\nu_\tau+\bar{\chi}\gamma^\mu\chi)Z'_\mu,
\end{equation}
where $g'$ is the new gauge coupling. For simplicity, the direct kinetic mixing term between the new and hypercharge gauge field is assumed to be zero. However, radiative corrections at the one-loop level still induce finite mixing as
\begin{equation}\label{Eq:Mix}
	\mathcal{L}\supset - \frac{\varepsilon}{2}F_{\mu\nu} F^{\prime \mu \nu},
\end{equation}
where $F_{\mu \nu}$ and $F^{\prime}_{\mu \nu}$ are the field strength for photon and $Z'$, respectively. The mixing parameter is calculated as \cite{Escudero:2019gzq}
\begin{eqnarray}
	\varepsilon = - \frac{e g'}{12\pi^2} \ln \frac{m_\tau^2}{m_\mu^2}\simeq -\frac{g'}{70}.
\end{eqnarray}

The dominant partial decay widths of $Z'$ are
\begin{eqnarray}
	\Gamma_{Z'\to f\bar{f}}= \frac{k_f g'^2 m_{Z'}}{12\pi} \left(1+\frac{2 m_f^2}{m_{Z'}^2}\right)\sqrt{1-\frac{4 m_f^2}{m_{Z'}^2}},
\end{eqnarray}
where $k_f=1$ for $f=\mu,\tau,\chi$ and $k_f=1/2$ for $f=\nu_\mu,\nu_\tau$. Through kinetic mixing, the partial width of $Z'\to e^+e^-$ is given by
\begin{equation}
	\Gamma_{Z'\to e^+e^-} = \frac{(\varepsilon e)^2 m_{Z'}}{12\pi} \left(1+\frac{2 m_e^2}{m_{Z'}^2}\right)\sqrt{1-\frac{4 m_e^2}{m_{Z'}^2}},
\end{equation}
which is typically suppressed by the mixing \cite{Bauer:2018onh}.

Contributions to  the muon magnetic moment of $Z'$ is at one-loop level, which is calculated as \cite{Baek:2001kca}
\begin{equation}
	\Delta a_\mu = \frac{g'^2}{4\pi^2}\int_0^1 dx \frac{m_\mu^2 x(1-x)^2}{m_\mu^2(1-x)^2+m_{Z'}^2 x^2}.
\end{equation}
To explain the anomaly $\Delta a_\mu=(251\pm59)\times10^{-11}$ \cite{Muong-2:2021ojo}, $g'\sim 5\times10^{-4}$ with $m_{Z'}\sim\mathcal{O}(10)$ MeV is required, which is now stringently constrained by the latest NA64$\mu$ experiment \cite{NA64:2024klw}.

\subsection{Constraints}\label{Sec:Cs}

\begin{figure}
	\begin{center}
		\includegraphics[width=0.8\linewidth]{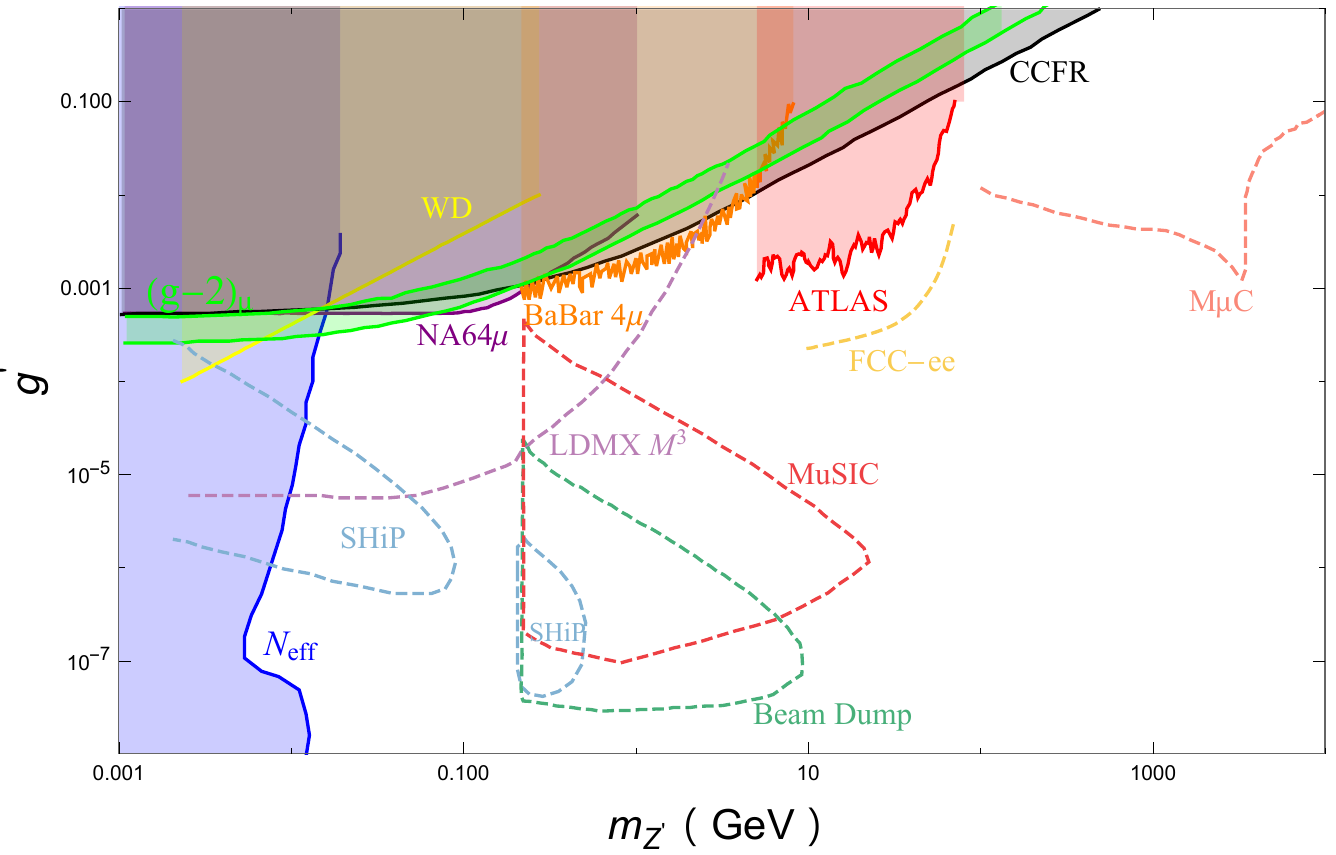}
	\end{center}
	\caption{Present and future constraints on the gauged $U(1)_{L_\mu-L_\tau}$. The green band is the $(g-2)_\mu$ favored region. The blue, yellow, purple, orange, red, and gray regions are excluded by $N_\text{eff}$ \cite{Escudero:2019gzq}, WD cooling \cite{Foldenauer:2024cdp}, NA64$\mu$ \cite{NA64:2024klw}, BaBar \cite{BaBar:2016sci}, ATLAS \cite{ATLAS:2024uvu}, and CCFR \cite{Altmannshofer:2014pba}, respectively. The dashed cyan, purple, brown, and pink lines are the future reach of SHiP \cite{Alekhin:2015byh}, LDMX \cite{Berlin:2018bsc}, FCC-ee \cite{Airen:2024iiy} and M$\mu$C \cite{Huang:2021nkl}. }
	\label{fig1}
\end{figure}

Decays of gauge boson $Z'$ will heat the neutrino population and delay the process of neutrino decoupling, which increases the effective number of neutrino species $N_\text{eff}$ \cite{Escudero:2019gzq}. The precise measurement of PLANCK result $N_\text{eff}=2.99\pm0.17$ \cite{Planck:2018vyg} is consistent with the standard model prediction $N_\text{eff}^\text{SM}=3.044$ \cite{Akita:2020szl,Froustey:2020mcq}, which disfavors the light $Z'$ region with $m_{Z'}\lesssim 10$ MeV \cite{Escudero:2019gzq}.

The gauge boson $Z'$ could affect the evolution of white dwarfs by enhancing the plasmon decay into neutrinos. Around $m_{Z'}\sim10$  MeV, the recent analysis in Ref.~\cite{Foldenauer:2024cdp} shows that the constraint from white dwarfs (WD) cooling is more stringent than the $\nu-e$ scattering by BOREXINO \cite{Gninenko:2020xys}.

The gauge boson $Z'$ also contributes to the neutrino trident production via the inelastic neutrino-nucleus scattering $\nu_\mu N\to \nu_\mu N \mu^+\mu^-$ \cite{Altmannshofer:2014pba}. The CCFR experiment has measured the normalized cross section $\sigma/\sigma_\text{SM}=0.82\pm0.28$ \cite{CCFR:1991lpl}, which has excluded the heavy $Z'$ region with $m_{Z'}\gtrsim0.3$~GeV for muon anomalous magnetic moment.

For light gauge boson below the dimuon threshold, $Z'\to \bar{\nu} \nu$ is the dominant decay mode \cite{Bauer:2018onh}. The gauge boson $Z'$ could be tested through the bremsstrahlung process $\mu N\to \mu N Z', Z' \to \bar{\nu}\nu $ by the high energy muon scattering off nucleus $N$, which leads to the signature of missing momentum in the final state. The region with $g'\gtrsim6\times10^{-4}$ and $m_{Z'}\sim \mathcal{O}(10)$ MeV is now excluded by the NA64$\mu$ experiment \cite{NA64:2024klw}. In the future, the NA64 \cite{Gninenko:2014pea} and LDMX \cite{Berlin:2018bsc} experiments are expected to probe $g'\gtrsim10^{-5}$ for $m_{Z'}\lesssim 1$ GeV. Meanwhile, the future SHiP experiment could probe $g'\gtrsim10^{-6}$ for $m_{Z'}\sim\mathcal{O}(10)$ MeV \cite{Alekhin:2015byh}.

Above the dimuon threshold, $Z'\to \mu^+\mu^-$ becomes the promising channel. The searches of the process $e^+e^-\to \mu^+\mu^- Z', Z'\to \mu^+\mu^-$ at BABAR \cite{BaBar:2016sci}  and Belle-II \cite{Belle-II:2024wtd} have excluded $g'\gtrsim10^{-3}$ for $Z'$ mass in the range of 0.212--10 GeV. Through the muon bremsstrahlung process $\mu~\text{Au}\to \mu ~\text{Au}~ Z'$, the proposed MuSIC experiment could probe the parameter space with $g'\gtrsim10^{-7}$ for $m_{Z'}\in[0.212,20]$ GeV \cite{Davoudiasl:2024fiz}. The future Beam Dump experiment could further reach $g'\sim4\times 10^{-8}$ \cite{Cesarotti:2023sje}. At hadron collider, searches for $Z'$ have been performed in the process as $pp\to Z^{(*)}\to Z' \mu^+\mu^-\to \mu^+\mu^-\mu^+\mu^- $ \cite{ATLAS:2023vxg} and $pp\to W^{\pm(*)}\to Z' \mu^\pm\nu\to \mu^+\mu^-\mu^\pm\nu $ \cite{ATLAS:2024uvu}, which have excluded $g'\gtrsim0.003$ in the $Z'$ mass range of 5--81 GeV. In the future, the FCC-ee collider could push this limit down to $g'\gtrsim10^{-4}$ \cite{Airen:2024iiy}. Meanwhile, the muon collider could probe a large parameter space for $Z'$ above the electroweak scale \cite{Huang:2021nkl,Dasgupta:2023zrh}.

In Figure \ref{fig1}, we summarize the latest and future constraints on gauged $U(1)_{L_\mu-L_\tau}$. Light $Z'$ below 10~MeV is now excluded by $N_\text{eff}$. The current experiments typically exclude $g'\gtrsim10^{-3}$ in the range of 10~MeV -- 100 GeV. For $m_{Z'}$ below 0.2 GeV, a large portion of the parameter space could be covered by the LDMX \cite{Berlin:2018bsc} and SHiP experiments \cite{Alekhin:2015byh}. For $m_{Z'}$ in the range of $0.2-20$ GeV, the MuSIC \cite{Davoudiasl:2024fiz} and Beam Dump experiments \cite{Cesarotti:2023sje} would test $g'\gtrsim4\times10^{-8}$. For $m_{Z'}\gtrsim\mathcal{O}(10)$ GeV, FCC-ee \cite{Airen:2024iiy} is the ideal machine to probe the new gauge boson. Above the electroweak scale, the future muon collider \cite{Huang:2021nkl,Dasgupta:2023zrh} could reveal most region with $g'\gtrsim 10^{-3}$.

\section{Dark Matter}\label{Sec:DM}

\subsection{Relic Density}
In this paper, we consider dark matter produced through the freeze-out mechanism.
The evolution of the dark matter relic density is then determined by the Boltzmann equation \cite{Holst:2021lzm}
\begin{equation}
	\frac{d n}{dt}+ 3 H n = - \frac{\langle \sigma v \rangle}{2} (n^2-n_\text{eq}^2),
\end{equation}
where $H$ is the Hubble parameter, $n=n_\chi+n_{\bar{\chi}}$ is the total number density of dark matter, and $n_\text{eq}$ is the equilibrium number density.

If kinematically allowed, the dark matter $\chi$ could annihilate via the $s$-channel processes $\chi\bar{\chi}\to f\bar{f}~(f=\mu,\tau,\nu_\mu,\nu_\tau)$ and the $t$-channel process $\chi\bar{\chi}\to Z'Z'$. The corresponding thermally averaged annihilation cross sections are
\begin{eqnarray}
	\langle \sigma v \rangle_{\chi \bar{\chi}\to f\bar{f}} &\simeq&  \frac{\kappa_f g'^4 }{2\pi m_\chi} \frac{(2m_\chi^2+m_f^2)(m_\chi^2-m_f^2)^{1/2}}{(4m_\chi^2-m_{Z'}^2)^2+m_{Z'}^2 \Gamma_{Z'}^2},\\
	\langle \sigma v \rangle_{\chi \bar{\chi}\to Z'Z'} &\simeq& \frac{g'^4}{4\pi m_\chi} \frac{(m_\chi^2-m_{Z'}^2)^{3/2}}{(2m_\chi^2-m_{Z'}^2)^2},
\end{eqnarray}
where $\Gamma_{Z'}$ is the total decay width of $Z'$. To properly calculate the annihilation cross section near the $Z'$ pole \cite{Griest:1990kh}, we use micrOMEGAs \cite{Belanger:2013oya} to obtain the numerical results of the relic density.

\begin{figure}
	\begin{center}
		\includegraphics[width=0.8\linewidth]{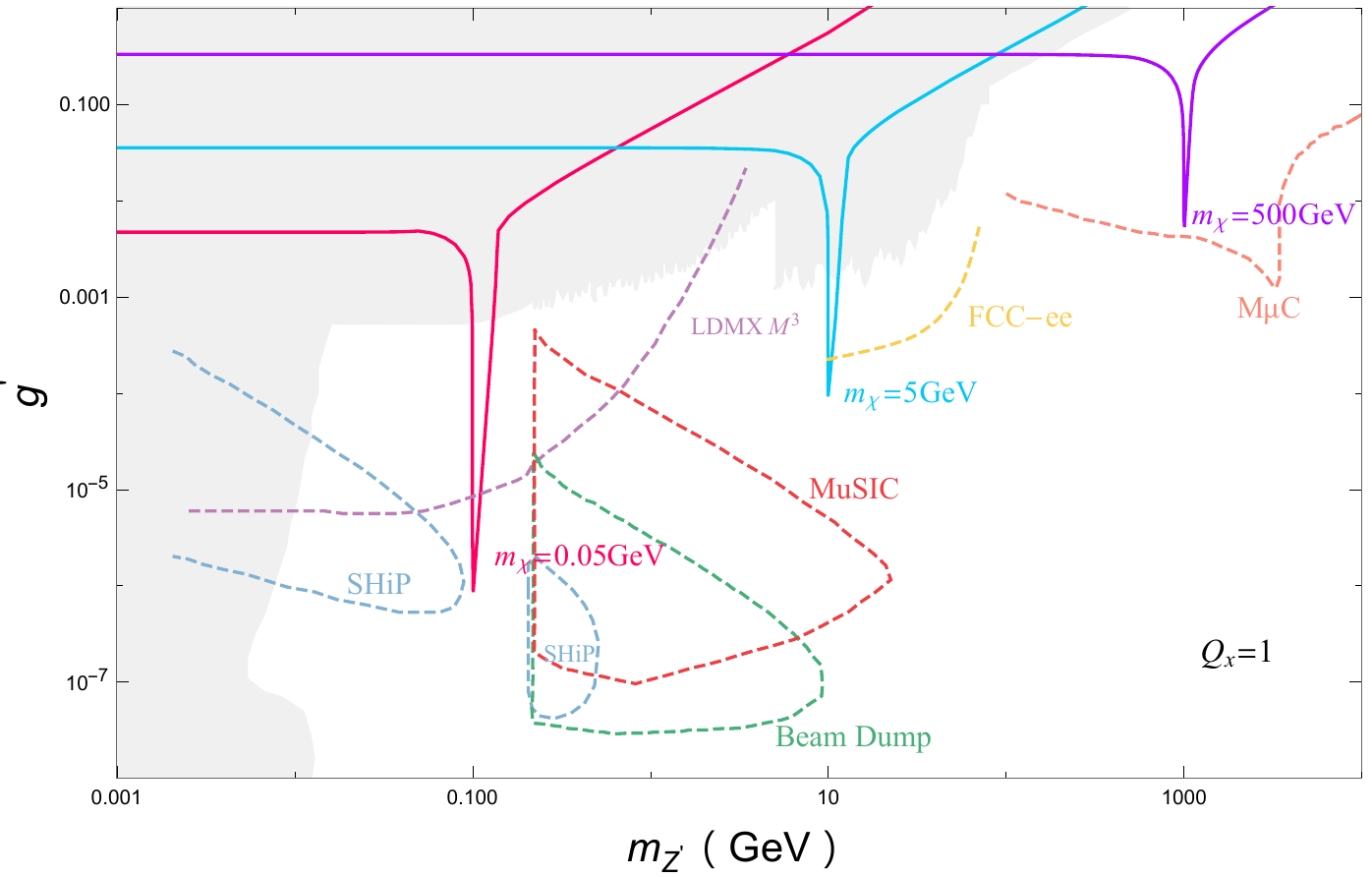}
	\end{center}
	\caption{ Required values of $g'$ and $m_{Z'}$ to generate the observed dark matter relic density. The red, blue, and purple lines correspond to dark matter mass $m_\chi=0.05,5,500$ GeV, respectively. The gray region is excluded by current searches. Other labels are the same as Figure~\ref{fig1}. }
	\label{fig2}
\end{figure}

In Figure \ref{fig2}, we show the required values of $g'$ and $m_{Z'}$ to generate the observed dark matter relic density $\Omega_\chi h^2=0.12\pm0.001$ \cite{Planck:2018vyg}. With stringent constraints, the sub-GeV and GeV scale dark matter can only satisfy relic density near the $Z'$ resonance. The future LDMX \cite{Berlin:2018bsc} and FCC-ee experiment \cite{Airen:2024iiy} would probe such scenarios. Meanwhile, the SHiP \cite{Alekhin:2015byh} and MuSIC experiment \cite{Davoudiasl:2024fiz} are expected to test certain resonance region. For dark matter mass larger than the electroweak scale, the $\chi\bar{\chi}\to Z'Z'$ channel could also have a significant contribution to relic density. It is also clear that the muon collider \cite{Huang:2021nkl,Dasgupta:2023zrh} is able to cover the whole $Z'$ resonance region around the TeV scale.

\subsection{Direct Detection}

\begin{figure}
	\begin{center}
		\includegraphics[width=0.8\linewidth]{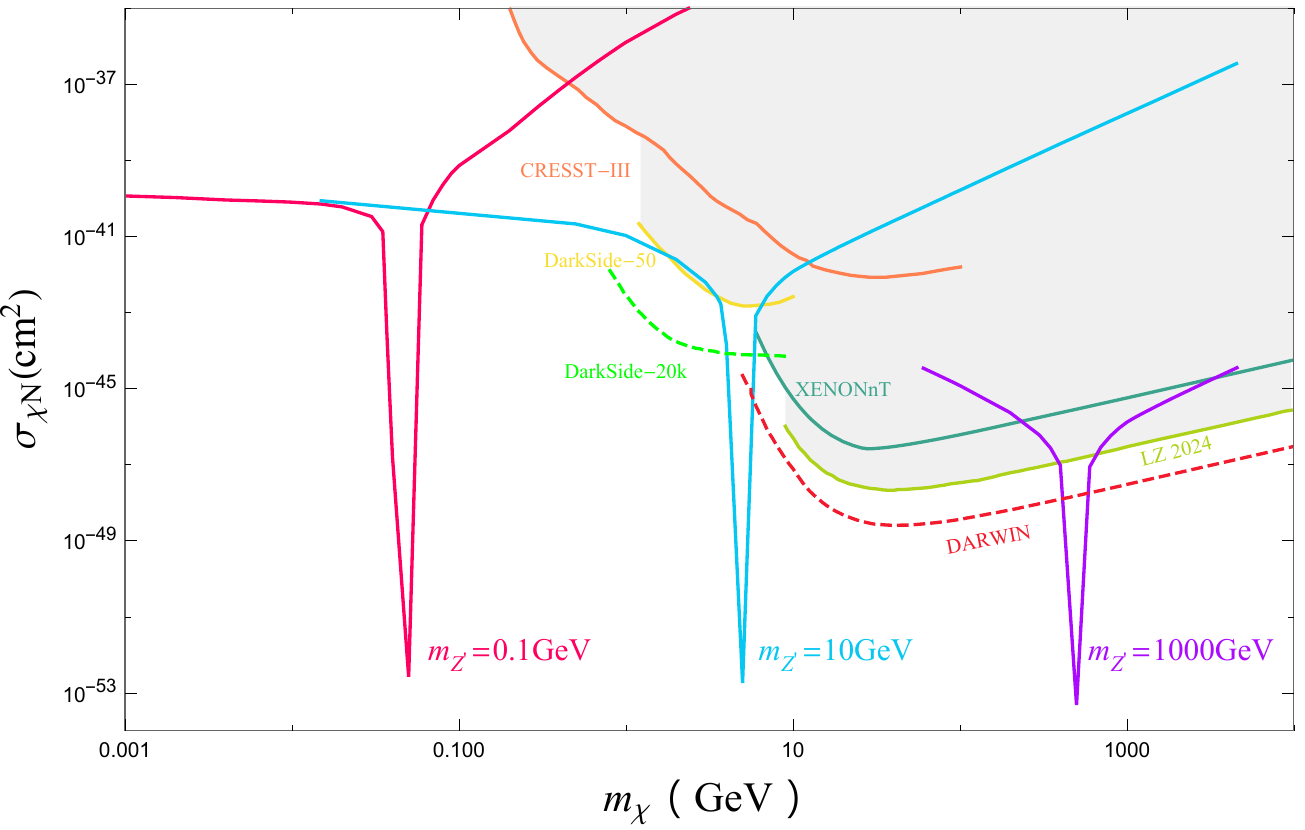}
	\end{center}
	\caption{Predicted DM-nucleon scattering cross section and various experimental limits. The red, blue, and purple lines are the benchmark scenario with $m_{Z'}=0.1,10$, and 1000 GeV, respectively. The orange, yellow, dark green, and light green solid lines are the limits of CRESST-III \cite{CRESST:2019jnq}, DarkSide-50 \cite{DarkSide-50:2022qzh}, XENONnT \cite{XENON:2023cxc}, and LZ \cite{LZCollaboration:2024lux}. The green and red dashed lines are the future reach of DarkSide-20k \cite{DarkSide-20k:2024yfq} and DARWIN \cite{DARWIN:2016hyl}.  }
	\label{fig3}
\end{figure}

Although the dark matter $\chi$ does not couple directly to quarks or electrons, the loop-induced kinetic mixing term in Eqn.~\eqref{Eq:Mix} contributes to the elastic DM-nucleon scattering. The spin-independent DM-nucleon scattering cross section is calculated as
\begin{equation}
	\sigma_{\chi N}=\frac{\mu_N^2}{\pi} \frac{Z^2 g'^2 \varepsilon^2 e^2}{A^2 m_{Z'}^4},
\end{equation}
where $\mu_N = m_\chi m_N/(m_\chi+m_N)$ is the reduced DM-nucleon mass. $A$ and $Z$ are the mass and atomic number of the nucleus.

Currently, the CRESST-III experiment \cite{CRESST:2019jnq} has excluded $\sigma_{\chi N}\gtrsim10^{-38}~ \text{cm}^2$ for sub-GeV DM. The DarkSide-50 \cite{DarkSide-50:2022qzh} could exclude $\sigma_{\chi N}\gtrsim10^{-43}~ \text{cm}^2$ for GeV DM. In the mass range of $6-10$ GeV, XENONnT \cite{XENON:2023cxc} has set the most stringent limit. Above 10 GeV, LZ could exclude $\sigma_{\chi N}\gtrsim2\times10^{-48}~ \text{cm}^2$. In the near future, the DarkSide-20k \cite{DarkSide-20k:2024yfq} and DARWIN experiments \cite{DARWIN:2016hyl} could push the limit down to the level of neutrino fog \cite{OHare:2021utq}.

In Figure \ref{fig3}, we show the predicted DM-nucleon scattering cross section for the benchmark scenario with $m_{Z'}=0.1,10,1000$ GeV, where correct relic density $\Omega_\chi h^2=0.12$ and $g'<1$ are further required. Due to too small recoil energy, the DM-nucleon experiments lose the sensitivity for light DM $m_\chi\ll m_N$. So there is no DM-nucleon limit for $m_\chi\lesssim0.1$ GeV. Although (sub-)GeV DM in the non-resonance region could lead to a smaller scattering cross section than the current experimental limit, the direct searches for $Z'$ already disfavor such scenario with correct relic density. The predicted scattering cross section in the resonance region could be as small as $10^{-53}~\text{cm}^2$, which is far beyond the future experimental reach. For the canonical electroweak scale DM, it is obvious that the current LZ \cite{LZCollaboration:2024lux} has excluded the non-resonance region. As no other experimental limit in this region at present, the direct detection experiments set the most stringent constrain. A positive signature is also expected on the future DARWIN operation \cite{DARWIN:2016hyl}.

For light dark matter below 1 GeV, the constraints from DM-electron scattering would be more stringent. The DM-electron scattering cross section is calculated as
\begin{equation}
	\sigma_{\chi e} = \frac{\mu_e^2}{\pi} \frac{ g'^2 \varepsilon^2 e^2}{ (m_{Z'}^2+\alpha^2m_e^2)^2},
\end{equation}
where $\mu_e= m_\chi m_e/(m_\chi+m_e)$ is the reduced DM-electron mass, and $\alpha$ is the fine structure constant.

\begin{figure}
	\begin{center}
		\includegraphics[width=0.8\linewidth]{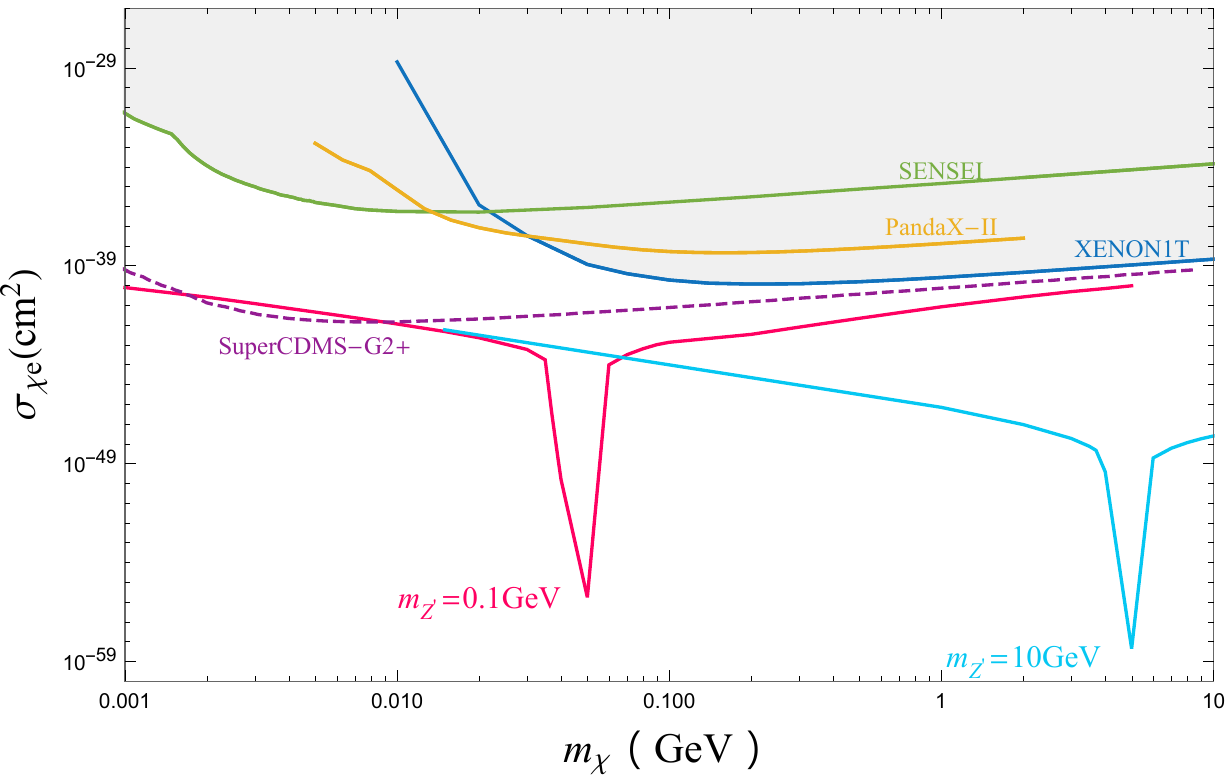}
	\end{center}
	\caption{ Predicted DM-electron scattering cross section and various experimental limits. The red and blue lines are the benchmark scenario with $m_{Z'}=0.1$ and 10 GeV. The green, orange, and dark blue lines are the limits of SENSEI \cite{SENSEI:2020dpa,SENSEI:2024yyt}, PandaX-II \cite{PandaX-II:2021nsg}, and XENON1T \cite{XENON:2019gfn}, respectively. The purple dashed line is the future limit from SuperCDMS-G2+ \cite{Battaglieri:2017aum}. }
	\label{fig4}
\end{figure}

Below 10 MeV, SENSEI \cite{SENSEI:2020dpa,SENSEI:2024yyt} could exclude $\sigma_{\chi e}\gtrsim10^{-36}~\text{cm}^2$. In the mass range of 10-30 MeV, PandaX-II \cite{PandaX-II:2021nsg} gives the most stringent limit. Above 30 MeV, XENON1T \cite{XENON:2019gfn} could exclude $\sigma_{\chi e}\gtrsim10^{-40}~\text{cm}^2$. In the future, SuperCDMS-G2+ \cite{Battaglieri:2017aum} would probe $\sigma_{\chi e}\gtrsim10^{-42}~\text{cm}^2$ for sub-GeV DM.

In Figure \ref{fig4}, we depict the predicted DM-electron scattering cross section with the correct relic density. Suppressed by the loop-induced kinetic mixing, we find that the theoretical predicted value is far smaller than the current experimental limits. In principle, the future SuperCDMS-G2+ experiment \cite{Battaglieri:2017aum} could probe the non-resonance region around $m_\chi\sim\mathcal{O}$(MeV). However, this region has been excluded by direct searches for $Z'$. Therefore, no positive signal is expected in the near future DM-electron experiments.

\subsection{Indirect Detection}

As shown in Figure \ref{fig2} and Figure \ref{fig3}, the combined results of relic density and direct detection experiments favor the $Z'$ resonance region. Considering the relatively loose constraints on DM annihilation into neutrinos \cite{Arguelles:2019ouk},  we also neglect the contribution from neutrino final states for simplicity, which could have a sizable impact for $m_\chi>100$ GeV \cite{Hambye:2021moy} in the following discussion. Therefore, we only consider the $s$-channel processes $\chi\bar{\chi}\to \ell^+\ell^-$ for the indirect detection constraints. 

\begin{figure}
	\begin{center}
		\includegraphics[width=0.8\linewidth]{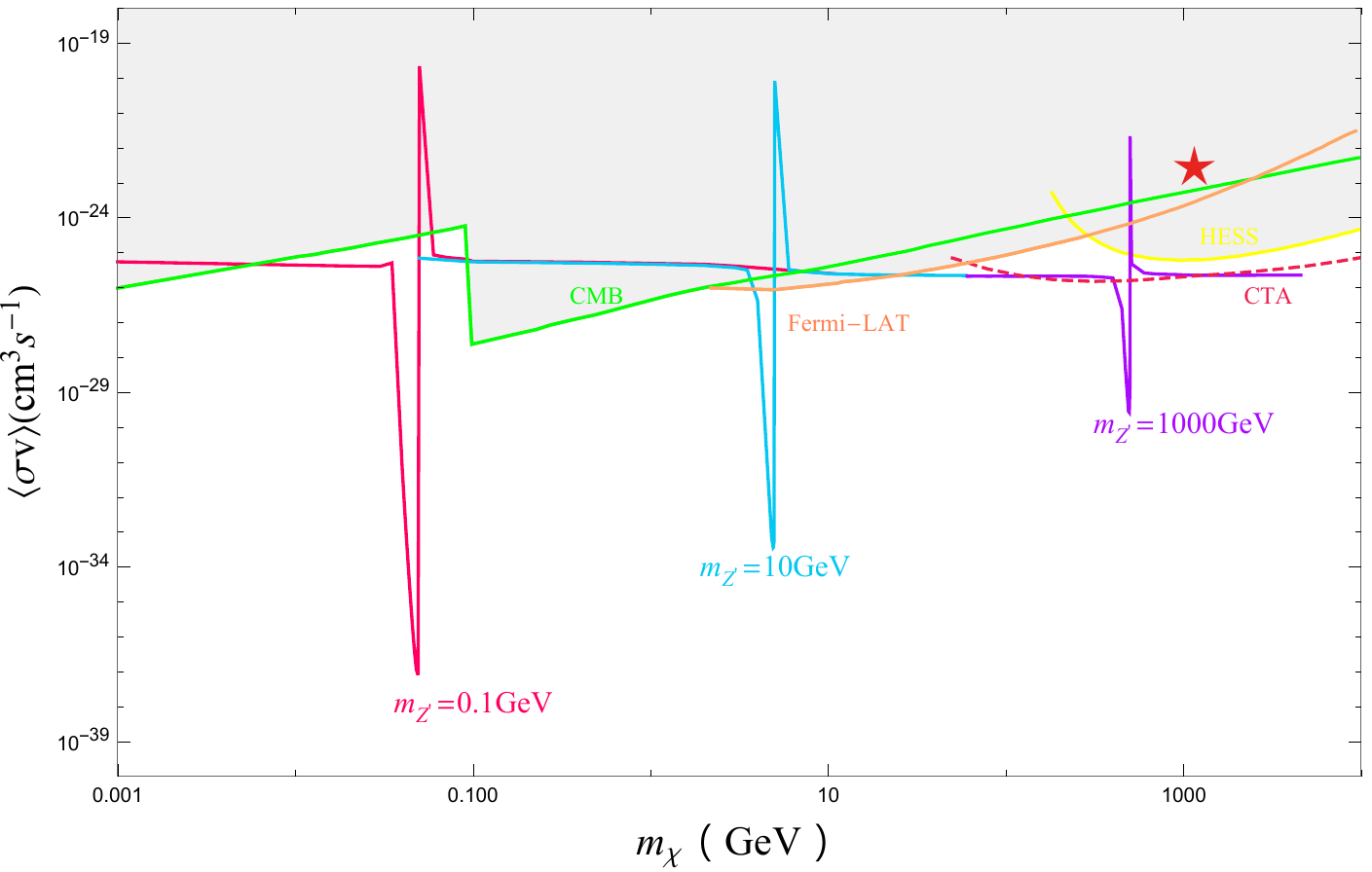}
	\end{center}
	\caption{ Predicted annihilation cross section and various constraints. The red, blue, and purple lines are the benchmark scenario with $m_{Z'}=$ 0.1,10, and 1000 GeV, respectively. The green, orange, and yellow lines are the current limit from CMB \cite{Dutta:2022wdi}, Fermi-LAT \cite{Fermi-LAT:2015att} and HESS \cite{HESS:2016mib}. The red dashed line is the future reach of the CTA experiment \cite{Pierre:2014tra}. The red star could interpret the AMS-02 observed positron excess \cite{AMS:2019rhg}.}
	\label{fig5}
\end{figure}

Annihilation of DM into charged fermions can inject energy to increase the ionization fraction during the era of recombination, which modifies the spectrum of CMB anisotropy \cite{Leane:2018kjk}. The Planck collaboration has set a stringent limit on the annihilation parameter \cite{Planck:2018vyg}
\begin{equation}
	p_\text{ann}=\sum_{\ell=e,\mu,\tau}\frac{1}{2} f_\text{eff}^{\ell}\cdot \text{FR}_\ell \frac{\langle \sigma v \rangle}{m_\chi}< 3.5\times10^{-28} \text{cm}^3 \text{s}^{-1} \text{GeV}^{-1},
\end{equation}
where $f_\text{eff}^\ell$ is the efficiency factor of the deposited energy for lepton $\ell$ \cite{Slatyer:2015jla}, FR$_{\ell}$ is the fraction of individual channel $\chi\bar{\chi}\to \ell^+\ell^-$ to the total DM annihilation cross section  $\langle \sigma v \rangle$. Different from previous results for individual channels of $e^+e^-,\mu^+\mu^-$, and $\tau^+\tau^-$ \cite{Dutta:2022wdi}, we obtain the CMB constraint on the total annihilation cross section $\langle \sigma v \rangle$ to compare with other limits. In this model, the fraction of $\chi\bar{\chi}\to e^+e^-$ is suppressed by the kinetic mixing \cite{Holst:2021lzm}. Therefore, the CMB constraint is relaxed when $m_\chi<m_\mu$ \cite{Foldenauer:2018zrz}.

The annihilation of DM could lead to high energy photons, which can be tested by Fermi-LAT \cite{Fermi-LAT:2015att} and HESS \cite{HESS:2016mib} experiments. The Fermi-LAT results could exclude 2 GeV $\lesssim m_\chi\lesssim100$ GeV for DM only annihilating into $\tau^+\tau^-$ final sate with the typical thermal cross section $\sim2\times10^{-26}~\text{cm}^3\text{s}^{-1}$ \cite{Fermi-LAT:2015att}. Around the TeV scale, the HESS experiment sets the most stringent limit, which could exclude $\langle\sigma v\rangle\gtrsim 2\times10^{-26}~\text{cm}^3\text{s}^{-1}$ in the pure $\tau^+\tau^-$ final state \cite{HESS:2016mib}. In the future, the CTA experiment would cover a large parameter space for DM above 100 GeV \cite{Pierre:2014tra}.

 In this model, the theoretical predicted differential photon flux is calculated as
\begin{equation}
	\frac{d\Phi_\gamma}{dE_\gamma}= \frac{\langle \sigma v \rangle}{8\pi m_\chi^2} \sum_{\ell=\mu,\tau} \text{FR}_\ell \frac{dN_\gamma^\ell}{dE_\gamma} \cdot J_\text{ann},
\end{equation}
where $E_\gamma$ is the energy of the photon, $dN^\ell_\gamma/dE_\gamma$ is the differential photon spectrum from lepton final states $\ell^+\ell^-$, and $J_\text{ann}$ is the astrophysical $J$-factor \cite{Fermi-LAT:2015att}. For $m_\chi\gtrsim10$ GeV,  the fraction is $\text{FR}_\mu\simeq\text{FR}_\tau\simeq 1/3$. Because the fraction of $e^+e^-$ final state is tiny for $U(1)_{L_\mu-L_\tau}$ DM in the indirect experiments sensitive region $m_\chi>2$ GeV, the contribution from $e^+e^-$ final sate is neglected. From the results of Fermi-LAT \cite{Fermi-LAT:2015att}, it is obvious that the constraint from the individual $\tau^+\tau^-$ final state is much more stringent than it from the individual $\mu^+\mu^-$ final state. Therefore, the photon flux in this model is dominant by the $\tau^+\tau^-$ final sate with equal fraction $\text{FR}_\mu\simeq\text{FR}_\tau\simeq 1/3$. Considering potential large astrophysical uncertainties, we then roughly estimate the indirect constraints by rescaling the experimental results as $\langle \sigma v \rangle\simeq \langle \sigma v \rangle_{\tau^+\tau^-}/\text{FR}_\tau\approx 3 \langle \sigma v \rangle_{\tau^+\tau^-}$, where $\langle \sigma v \rangle_{\tau^+\tau^-}$ is the indirect constraint for the individual $\tau^+\tau^-$ channel. More tedious but precise results can be obtained by performing the likelihood analysis \cite{Geringer-Sameth:2011wse}.

Figure \ref{fig5} shows the predicted annihilation cross section and corresponding constraints. In the non-resonance region, the typical annihilation cross section is $(2\sim3)\times10^{-26}~\text{cm}^3\text{s}^{-1}$. Current indirect detection then excludes the mass range of 0.1 -- 100 GeV. At the TeV scale, the HESS limit \cite{HESS:2016mib} is slightly larger than the thermal target of the non-resonance scenario. In the future, the CTA experiment \cite{Pierre:2014tra} is hopeful to probe such a region. In the resonance region, the annihilation cross section heavily depends on the mass relation between $Z'$ and $\chi$, due to different DM velocities at the time of freeze-out $v_f\sim\mathcal{O}(0.1)$ and at present $v_0\sim\mathcal{O}(0.001)$. When $m_\chi\simeq m_{Z'}/2$ is at the $Z'$ pole, the annihilation cross section can be greatly enhanced at present \cite{Ibe:2008ye}, which is disfavor by current experimental limits. On the other hand, when the pole condition $m_\chi\simeq m_{Z'}/(2\sqrt{1+v_f^2/\sqrt{2}})\sim0.49 m_{Z'}$ is satisfied at freeze-out, the present cross section is suppressed \cite{Guo:2009aj}. Therefore, $m_\chi$ slightly below the $Z'$ pole can easily avoid the tight constraints from indirect detection.

\begin{figure}
	\begin{center}
		\includegraphics[width=0.8\linewidth]{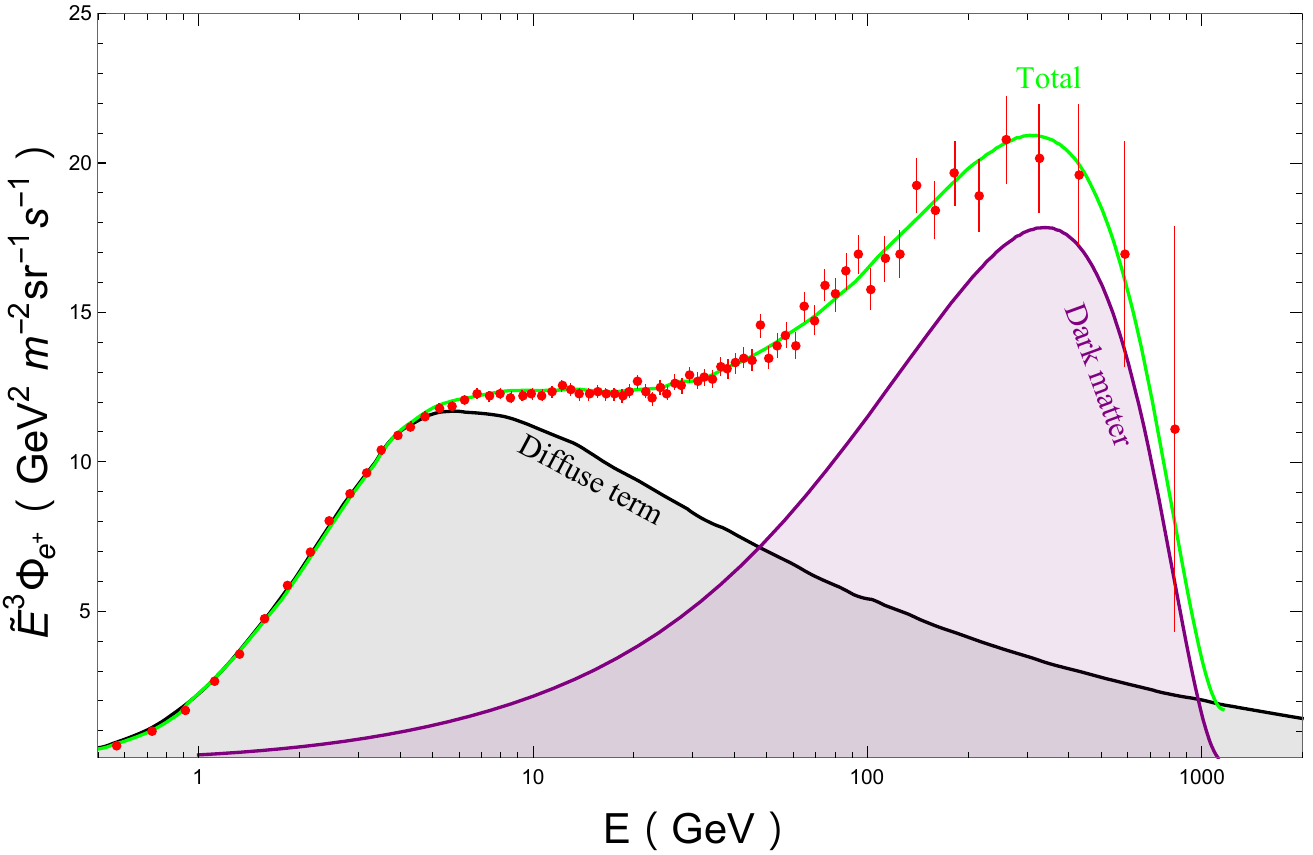}
	\end{center}
	\caption{ The flux of the cosmic positron measured by AMS-02 (red samples) \cite{AMS:2019rhg} and the theoretical total flux from dark matter annihilation (green line). The diffuse term contribution is the black line, and the source term from dark matter is the purple line. Here, $m_{Z'}/2\simeq m_\chi=1160$ GeV.}
	\label{fig6}
\end{figure}

Besides photons, indirect detection experiments also measure other cosmic rays, such as positrons. The AMS-02 collaboration has reported an excess of positron flux with a peak of around 300 GeV \cite{AMS:2019rhg}, which might be explained by dark matter in $U(1)_{L_\mu-L_\tau}$ \cite{He:2009ra}. In Figure \ref{fig6}, we show the benchmark point with $m_{Z'}/2\simeq m_\chi=1160$ GeV that could generate the observed positron excess. The induced positron flux from DM annihilation is obtained with  micrOMEGAs \cite{Belanger:2013oya}, where both the $\mu^+\mu^-$ and $\tau^+\tau^-$ final states are considered. It should be noted that the positron excess required annihilation cross section is about $\langle \sigma v \rangle\simeq 4\times 10^{-23}~\text{cm}^3 \text{s}^{-1}$. Although such large annihilation can be realized at the $Z'$ pole, all the indirect detection limits disfavor such a large annihilation cross section as shown in Figure \ref{fig5}. By fine tuning the mass relation of $m_\chi$ and $m_{Z'}$, the indirect detection limits on DM for positron excess might be relaxed due to velocity dependent annihilation \cite{Xiang:2017jou}. Anyway, the positron excess requires TeV scale $Z'$ with $g'\sim\mathcal{O}(0.01)$, which is within the reach of future muon collider \cite{Huang:2021nkl,Dasgupta:2023zrh}.

\subsection{Combined Result}

\begin{figure}
	\begin{center}
		\includegraphics[width=0.8\linewidth]{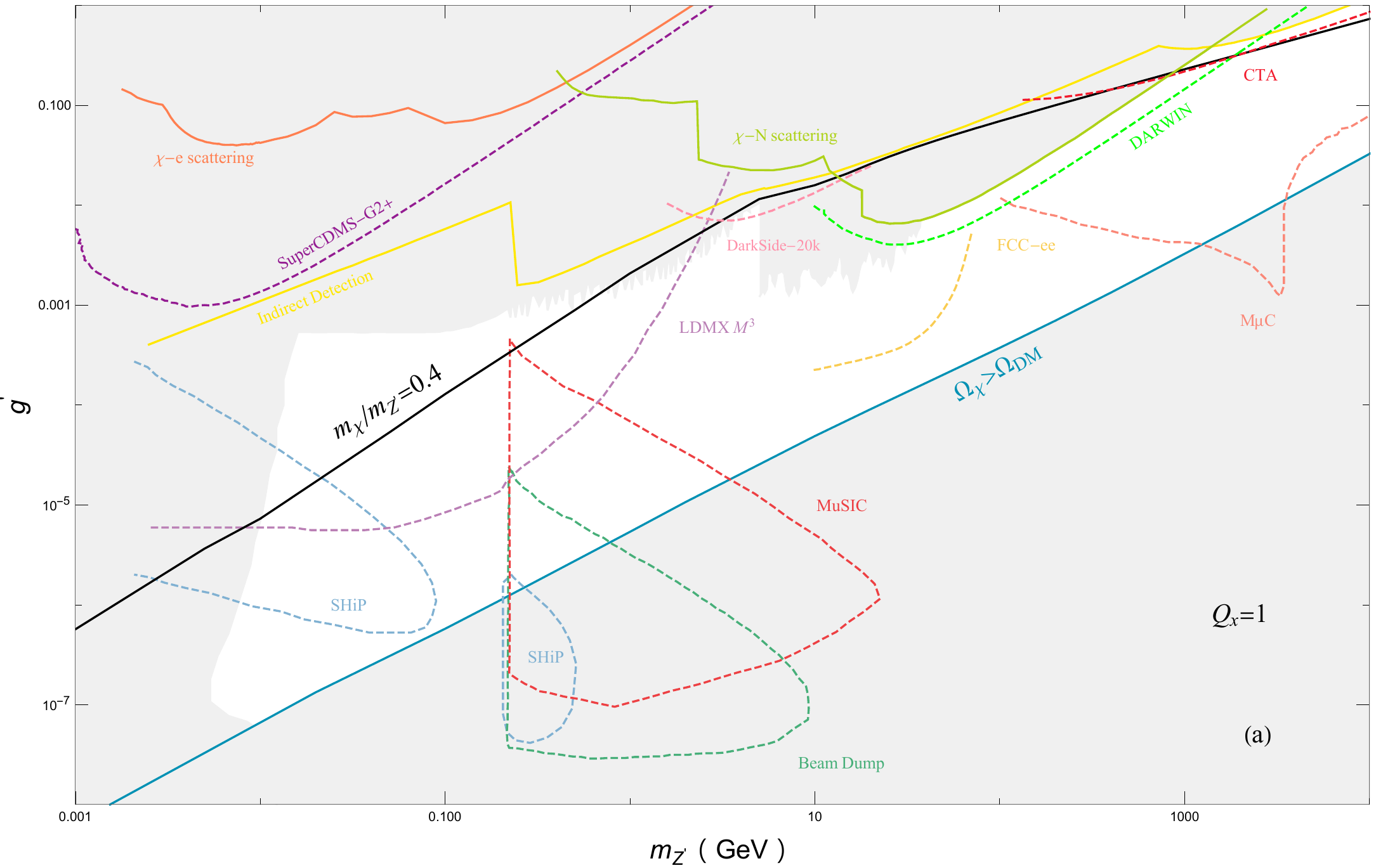}
		\includegraphics[width=0.8\linewidth]{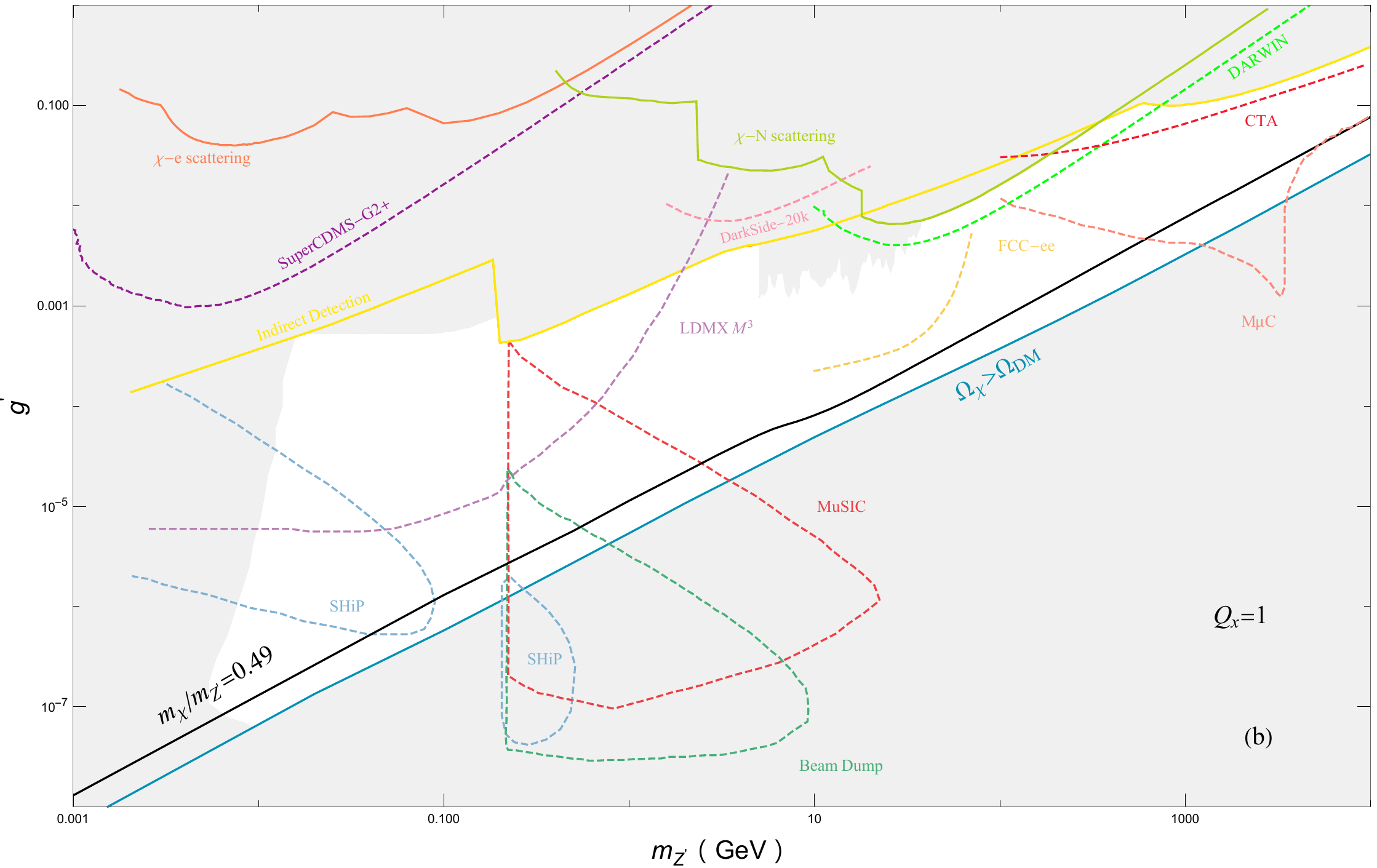}
	\end{center}
	\caption{ The combined parameter space of benchmark scenario $m_\chi/m_{Z'}=0.4$ in panel (a) and  $m_\chi/m_{Z'}=0.49$ in panel (b). The orange, green, and yellow solid lines are the combined limit from $\chi-e$ scattering, $\chi-N$ scattering, and indirect detection at present, respectively. The black lines are the required values of $g'$ and $m_{Z'}$ for correct relic density. Below the blue lines, the DM relic density is over abundance even near the $Z'$ pole for $Q_\chi=1$.  }
	\label{fig7}
\end{figure}

The experimental searches for new gauge boson $Z'$ have set stringent constraints for light $Z'$ below about 100 GeV. Meanwhile, the direct and indirect detection experiments on dark matter mainly constrain heavy DM above 10 GeV. In this section, we combine all the relevant constraints at present and in the future. As the conventional benchmark scenario with $m_\chi/m_{Z'}=1/3$ is already disfavored by various experiments \cite{Foldenauer:2018zrz}, we consider two new benchmark scenarios in this paper: one is $m_\chi/m_{Z'}=0.4$, and the other one is $m_\chi/m_{Z'}=0.49$. The results are shown in Figure \ref{fig7}.

It is obvious that the current DM-electron scattering experiments can only probe $g'\gtrsim0.1$ and $m_{Z'}\lesssim1$~GeV. Although the future SuperCDMS-G2+  \cite{Battaglieri:2017aum} could test $g'\gtrsim0.001$, such region is already excluded by current direct searches of $Z'$. Therefore, no positive signature is expected in the future DM-electron experiments for the $Z'$ portal DM $U(1)_{L_\mu-L_\tau}$. On the other hand, the limits from DM-nucleon scattering and indirect detection become the most stringent constraints for $m_{Z'}\gtrsim100$ GeV, as no direct search of $Z'$ in such a region. In the future, the muon collider \cite{Huang:2021nkl,Dasgupta:2023zrh} would cover most of the parameter space for correct relic density.

For the benchmark scenario with $m_\chi/m_{Z'}=0.4$, the viable region is separated into two parts. The light sub-GeV region is totally within the reach of LDMX \cite{Berlin:2018bsc}, and could be further confirmed by SHiP for $m_{Z'}\sim\mathcal{O}(0.01)$ GeV. However, the MuSIC \cite{Davoudiasl:2024fiz} and Beam Dump experiments \cite{Cesarotti:2023sje} can hardly probe such a large $g'$ required by relic density. The GeV to TeV mass range set by relic density is now excluded by direct search of $Z'$ and direct detection. Around $m_{Z'}\approx 2$ TeV, the predicted DM-nucleon scattering cross section can be tested at DARWIN \cite{DARWIN:2016hyl}, and the indirect detection CTA experiment \cite{Pierre:2014tra} is also capable of probing such a region. The typical relic density required gauge coupling $g'\sim\mathcal{O}(0.1)$ can be easily confirmed at the muon collider \cite{Huang:2021nkl,Dasgupta:2023zrh}. Therefore, this region with $m_{Z'}\approx 2$ TeV is the most promising one in the forthcoming experiments.

Different from the previous benchmark scenario with $m_\chi/m_{Z'}=0.4$, the benchmark scenario with $m_\chi/m_{Z'}=0.49$ is close to the $Z'$ resonance. As the annihilation cross section at the time of freeze-out is enhanced by the Breit–Wigner propagator, the relic density required gauge coupling for $m_\chi/m_{Z'}=0.49$  is typically two orders of magnitudes smaller than it for  $m_\chi/m_{Z'}=0.4$. In this way, we find that $m_{Z'}\gtrsim0.01$~GeV is all viable at present. Although the suppressed gauge coupling makes this scenario satisfy all constraints, it also leads this scenario hard to detect. For instance, all the dark matter detection experiments can not probe this scenario. The SHiP experiment  \cite{Alekhin:2015byh} could probe $m_{Z'}\simeq0.08$ GeV, while the MuSIC experiment \cite{Davoudiasl:2024fiz} could test $Z'$ in the mass range of 0.2--2 GeV. The TeV muon collider \cite{Huang:2021nkl,Dasgupta:2023zrh} would explore the region with $m_{Z'}\gtrsim600$ GeV. To cover the parameter space around 100 GeV, a muon collider operating at the electroweak scale is strongly encouraged.

\section{Conclusion}\label{Sec:CL}

Although the new gauge boson $Z'$ in $U(1)_{L_\mu-L_\tau}$ is extensively studied  as the interpretation of muon $g-2$, the latest experimental result does not favor the corresponding parameter space. In this paper, we revisit the minimal $Z'$ portal dark matter in the gauged $U(1)_{L_\mu-L_\tau}$ model, which introduces a singlet Dirac fermion $\chi$ with the $U(1)_{L_\mu-L_\tau}$ charge $Q_\chi=1$ as the dark matter candidate.

The combined analysis shows that light dark matter suffers stringent constraints from searches for $Z'$. For example, $m_{Z'}$ less than 10 MeV is excluded by the effective number of neutrino species $N_\text{eff}$. And the new gauge coupling $g'$ should be smaller than 0.001 to satisfy current limits. On the other hand, the direct and indirect detection experiments set the most strict constraints for the heavy dark matter. To satisfy current experimental limits, dark matter near the $Z'$ resonance region is favored.

In the $U(1)_{L_\mu-L_\tau}$ model, the coupling of $Z'$ to the electron is suppressed by the kinetic mixing term. Under current limits on $Z'$, no positive signature is expected in the future DM-electron experiments. The most promising scenario is $m_{Z'}\approx 2$ TeV with $m_\chi/m_{Z'}\simeq 0.4$, which would lead to observable signatures at the direct, indirect, and collider. Increasing the mass ratio $m_\chi/m_{Z'}$ makes the model easier to escape current limits, but also becomes harder to detect in dark matter experiments. Meanwhile, various experiments are proposed to test the new gauge boson $Z'$, which covers the most viable parameter space of the dark matter in the future.

\section*{Acknowledgments}
This work is supported by the National Natural Science Foundation of China under Grant No. 11805081, Natural Science Foundation of Shandong Province under Grant No. ZR2022MA056 and ZR2024QA138, the Open Project of Guangxi Key Laboratory of Nuclear Physics and Nuclear Technology under Grant No. NLK2021-07, University of Jinan Disciplinary Cross-Convergence Construction Project 2024 (XKJC-202404).


\end{document}